\documentclass{bmcart}

\usepackage{amsthm,amsmath}
\usepackage{mathrsfs}
\usepackage[utf8]{inputenc} %unicode support
\usepackage{graphicx}
\usepackage{float}
\usepackage{todonotes}
\usepackage{multirow}

\startlocaldefs
\endlocaldefs

\begin{document}

%%% Start of article front matter
\begin{frontmatter}

\begin{fmbox}
\dochead{Research}

\title{GenHap: A Novel Computational Method Based on Genetic Algorithms for Haplotype Assembly}

%%%%%%%%%%%%%%%%%%%%%%%%%%%%%%%%%%%%%%%%%%%%%%
%%                                          %%
%% Enter the authors here                   %%
%%                                          %%
%% Specify information, if available,       %%
%% in the form:                             %%
%%   <key>={<id1>,<id2>}                    %%
%%   <key>=                                 %%
%% Comment or delete the keys which are     %%
%% not used. Repeat \author command as much %%
%% as required.                             %%
%%                                          %%
%%%%%%%%%%%%%%%%%%%%%%%%%%%%%%%%%%%%%%%%%%%%%%

\author[
   addressref={aff1},% id's of addresses, e.g. {aff1,aff2}
   corref={aff1},% id of corresponding address, if any
%    noteref={n1},% id's of article notes, if any
   email={andrea.tangherloni@disco.unimib.it} % email address
]{\inits{AT}\fnm{Andrea} \snm{Tangherloni}}
\author[
   addressref={aff1},
   email={simone.spolaor@disco.unimib.it}
]{\inits{SS}\fnm{Simone} \snm{Spolaor}}
\author[
   addressref={aff1,aff5},
   email={leonardo.rundo@disco.unimib.it}
]{\inits{LR}\fnm{Leonardo} \snm{Rundo}}
\author[
   addressref={aff1,aff6},
   email={nobile@disco.unimib.it},
]{\inits{MSN}\fnm{Marco S.} \snm{Nobile}}
\author[
   addressref={aff2,aff6},
   email={paolo.cazzaniga@unibg.it}
]{\inits{PC}\fnm{Paolo} \snm{Cazzaniga}}
\author[
   addressref={aff1,aff6},
   email={mauri@disco.unimib.it}
]{\inits{GM}\fnm{Giancarlo} \snm{Mauri}}
\author[
   addressref={aff3},
   email={pl219@cam.ac.uk}
]{\inits{PL}\fnm{Pietro} \snm{Liò}}
\author[
   addressref={aff4},
   email={ivan.merelli@itb.cnr.it}
]{\inits{IV}\fnm{Ivan} \snm{Merelli}}
\author[
   addressref={aff1},
   email={daniela.besozzi@unimib.it}
]{\inits{DB}\fnm{Daniela} \snm{Besozzi}}

%%%%%%%%%%%%%%%%%%%%%%%%%%%%%%%%%%%%%%%%%%%%%%
%%                                          %%
%% Enter the authors' addresses here        %%
%%                                          %%
%% Repeat \address commands as much as      %%
%% required.                                %%
%%                                          %%
%%%%%%%%%%%%%%%%%%%%%%%%%%%%%%%%%%%%%%%%%%%%%%

\address[id=aff1]{%% unique id
  \orgname{Department of Informatics, Systems and Communication, University of Milano-Bicocca}, % university, etc
  \street{Viale Sarca 336}, %
  \postcode{20126} % post or zip code
  \city{Milano}, % city
  \cny{Italy}% country
}
\address[id=aff2]{
  \orgname{Department of Human and Social Sciences, University of Bergamo},
  \street{Piazzale Sant'Agostino 2},
  \postcode{24129}
  \city{Bergamo},
  \cny{Italy}
}
\address[id=aff3]{
  \orgname{Computer Laboratory, University of Cambridge},
  \street{15 JJ Thomson Avenue},
  \postcode{CB3 0FD}
  \city{Cambridge},
  \cny{UK}
}
\address[id=aff4]{
  \orgname{Institute of Biomedical Technologies, Italian National Research Council},
  \street{Via Fratelli Cervi 93},
  \postcode{20090}
  \city{Segrate (MI)},
  \cny{Italy}
}
\address[id=aff5]{
  \orgname{Institute of Molecular Bioimaging and Physiology, Italian National Research Council},
  \street{Contrada Pietrapollastra-Pisciotto},
  \postcode{90015}
  \city{Cefalù (PA)},
  \cny{Italy}
}
\address[id=aff6]{
	\orgname{SYSBIO.IT Centre of Systems Biology},
	\street{Piazza della Scienza 2},
	\postcode{20126}
	\city{Milano},
	\cny{Italy}
}

%%%%%%%%%%%%%%%%%%%%%%%%%%%%%%%%%%%%%%%%%%%%%%
%%                                          %%
%% Enter short notes here                   %%
%%                                          %%
%% Short notes will be after addresses      %%
%% on first page.                           %%
%%                                          %%
%%%%%%%%%%%%%%%%%%%%%%%%%%%%%%%%%%%%%%%%%%%%%%

\begin{artnotes}
%\note{Sample of title note}     % note to the article
% \note[id=n1]{Equal contributor} % note, connected to author
\end{artnotes}

\end{fmbox}% comment this for two column layout

%%%%%%%%%%%%%%%%%%%%%%%%%%%%%%%%%%%%%%%%%%%%%%
%%                                          %%
%% The Abstract begins here                 %%
%%                                          %%
%% Please refer to the Instructions for     %%
%% authors on http://www.biomedcentral.com  %%
%% and include the section headings         %%
%% accordingly for your article type.       %%
%%                                          %%
%%%%%%%%%%%%%%%%%%%%%%%%%%%%%%%%%%%%%%%%%%%%%%

\begin{abstractbox}

\begin{abstract} % abstract
\parttitle{Background}
In order to fully characterize the genome of an individual, the reconstruction of the two distinct copies of each chromosome, called haplotypes, is essential. 
The computational problem of inferring the full haplotype of a cell starting from read sequencing data is known as haplotype assembly, and consists in assigning all heterozygous Single Nucleotide Polymorphisms (SNPs) to exactly one of the two chromosomes.
Indeed, the knowledge of complete haplotypes is generally more informative than analyzing single SNPs and plays a fundamental role in many medical applications.

\parttitle{Results}
To reconstruct the two haplotypes, we addressed the weighted Minimum Error Correction (wMEC) problem, which is a successful approach for haplotype assembly.
This NP-hard problem consists in computing the two haplotypes that partition the sequencing reads into two disjoint sub-sets, with the least number of corrections to the SNP values.
To this aim, we propose here GenHap, a novel computational method for haplotype assembly based on Genetic Algorithms, yielding optimal solutions by means of a global search process.
In order to evaluate the effectiveness of our approach, we run GenHap on two synthetic (yet realistic) datasets, based on the Roche/454 and PacBio RS II sequencing technologies.
We compared the performance of GenHap against HapCol, an efficient state-of-the-art algorithm for haplotype phasing. 
Our results show that GenHap always obtains high accuracy solutions (in terms of haplotype error rate), and is up to $4\times$ faster than HapCol in the case of Roche/454 instances and up to $20\times$ faster when compared on the PacBio RS II dataset.
Finally, we assessed the performance of GenHap on two different real datasets.

\parttitle{Conclusions}
Future-generation sequencing technologies, producing longer reads with higher coverage, can highly benefit from GenHap, thanks to its capability of efficiently solving large instances of the haplotype assembly problem.
Moreover, the optimization approach proposed in GenHap can be extended to the study of allele-specific genomic features, such as expression, methylation and chromatin conformation, by exploiting multi-objective optimization techniques.
\end{abstract}

%%%%%%%%%%%%%%%%%%%%%%%%%%%%%%%%%%%%%%%%%%%%%%
%%                                          %%
%% The keywords begin here                  %%
%%                                          %%
%% Put each keyword in separate \kwd{}.     %%
%%                                          %%
%%%%%%%%%%%%%%%%%%%%%%%%%%%%%%%%%%%%%%%%%%%%%%


% MSC classifications codes, if any
%\begin{keyword}[class=AMS]
%\kwd[Primary ]{}
%\kwd{}
%\kwd[; secondary ]{}
%\end{keyword}

\end{abstractbox}
%
%\end{fmbox}% uncomment this for twcolumn layout

\end{frontmatter}

%%%%%%%%%%%%%%%%%%%%%%%%%%%%%%%%%%%%%%%%%%%%%%
%%                                          %%
%% The Main Body begins here                %%
%%                                          %%
%% Please refer to the instructions for     %%
%% authors on:                              %%
%% http://www.biomedcentral.com/info/authors%%
%% and include the section headings         %%
%% accordingly for your article type.       %%
%%                                          %%
%% See the Results and Discussion section   %%
%% for details on how to create sub-sections%%
%%                                          %%
%% use \cite{...} to cite references        %%
%%  \cite{koon} and                         %%
%%  \cite{oreg,khar,zvai,xjon,schn,pond}    %%
%%  \nocite{smith,marg,hunn,advi,koha,mouse}%%
%%                                          %%
%%%%%%%%%%%%%%%%%%%%%%%%%%%%%%%%%%%%%%%%%%%%%%

%%%%%%%%%%%%%%%%%%%%%%%%% start of article main body
% <put your article body there>

%%%%%%%%%%%%%%%%
%% Background %%
%%
\section*{Background}
Somatic human cells are diploids, that is, they contain 22 pairs of homologous chromosomes and a pair of sex chromosomes, one copy inherited from each parent.
In order to fully characterize the genome of an individual, the reconstruction of the two distinct copies of each chromosome, called haplotypes, is essential \cite{levy2007}.
The process of inferring the full haplotype information related to a cell is known as haplotyping, which consists in assigning all heterozygous Single Nucleotide Polymorphisms (SNPs) to exactly one of the two chromosome copies.
SNPs are one of the most studied genetic variations, since they play a fundamental role in many medical applications, such as drug-design or disease susceptibility studies, as well as in characterizing the effects of SNPs on the expression of phenotypic traits \cite{hirschhorn2005}.
This information can be valuable in several contexts, including linkage analysis, association studies, population genetics and clinical genetics \cite{snyder2015}.
Obviously, the complete set of SNPs of an individual (i.e., his/her haplotypes) is generally more informative than analyzing single SNPs, especially in the study of complex disease susceptibility.

Since a direct experimental reconstruction of haplotypes still requires huge sequencing efforts and is not cost-effective \cite{kuleshov2014}, computational approaches are extensively used to solve this problem.
In particular, two classes of methods exist for haplotype phasing \cite{snyder2015}.
The first class consists of statistical methods that try to infer haplotypes from genotypes sampled in a population.
These data, combined with datasets describing the frequency by which the SNPs are usually correlated in different populations, can be used to reconstruct the haplotypes of an individual. 
%The second class of methods analyzes sequencing read data, exploiting an idea similar to genome assembly.
%Their goal is to divide the entire set of reads into $k$ partitions, corresponding to the $k$ different haplotypes.
%In the case of diploid organisms, there are $2^n$ candidate haplotypes for $n$ \emph{heterozygous} SNP positions.
The second class of methods directly leverages sequencing data: in this case, the main goal is to partition the entire set of reads into two sub-sets, exploiting the partial overlap among them in order to ultimately reconstruct the corresponding two different haplotypes of a diploid organism \cite{patterson2015}.
The effectiveness of these methods was limited by the length of the reads produced by second-generation sequencing technologies, which might be not long enough to span over a relevant number of SNP positions. 
This results in the reconstruction of short haplotype blocks \cite{zhang2002, daly2001}, since reads do not cover adjacent SNP positions adequately, hindering the possibility of reconstructing the full haplotypes.
However, in recent years the development of new sequencing technologies paved the way to the advent of the third-generation of sequencing platforms, namely PacBio RS II (Pacific Biosciences of California Inc., Menlo Park, CA, USA) \cite{rhoads2015,roberts2013} and Oxford Nanopore MinION (Oxford Nanopore Ltd., Oxford, United Kingdom) \cite{jain2015}, which are able to produce reads covering several hundreds of kilobases and spanning different SNP loci at once.
Unfortunately, the increased length comes at the cost of a decreased accuracy with respect to short and precise second-generation sequencing technologies, like NovaSeq (Illumina Inc., San Diego, CA, USA) \cite{quail2008}; thus, in order to obtain reliable data, the read coverage should be increased.

Among the computational methods for haplotype assembly, the Minimum Error Correction (MEC) is one of the most successful approaches.
This problem consists in computing the two haplotypes that partition the sequencing reads into two disjoint sets with the least number of corrections to the SNP values \cite{wang2005}.
Unfortunately, MEC was proven to be NP-hard \cite{lippert2002}.
A weighted variant of MEC, named weighted MEC (wMEC), was then proposed in \cite{greenberg2004}: the weights represent the confidence for the presence of a sequencing error, while the correction process takes into account the weight associated with each SNP value of a read.
These error schemes generally regard phred-scaled error probabilities and are very valuable for processing long reads generated by third-generation sequencing technologies, as they are prone to high sequencing error rates \cite{patterson2015}.

Several assembly approaches have been already proposed in literature.
Due to the NP-hardness of the MEC problem, some methods exploit heuristic strategies.
Two noteworthy approaches are ReFHap \cite{duitama2010}, which is based on a heuristic algorithm for the Max-Cut problem on graphs, and ProbHap \cite{kuleshov2014ProbHap}, which generalizes the MEC formulation by means of a probabilistic framework.
In \cite{wang2005}, Wang \textit{et al.}  proposed a meta-heuristic approach based on Genetic Algorithms (GAs) to address an extended version of the MEC problem, called MEC with Genotype Information (MEC/GI), which also considers genotyping data during the SNP correction process.
A similar work was presented in \cite{wang2012}, where GAs are used to solve the MEC problem by using a fitness function based on a majority rule that takes into account the allele frequencies.
The results shown in \cite{wang2012} are limited to a coverage up to $10\times$ and a haplotype length equal to $700$.
More recently, an evolutionary approach called Probabilistic Evolutionary Algorithm with Toggling for Haplotyping (PEATH) was proposed in \cite{na2018}.
PEATH is based on the Estimation of Distribution Algorithm (EDA), which uses the promising individuals to build probabilistic models that are sampled to explore the search space.
This meta-heuristic deals with noisy sequencing reads, reconstructing the haplotypes under the all-heterozygous assumption.
These algorithms present some limitations, as in the case of ReFHap \cite{duitama2010}, ProbHap \cite{kuleshov2014ProbHap} and PEATH \cite{na2018}, which assume that the columns in the input matrix correspond to heterozygous sites \cite{chen2013}.
However, this all-heterozygous assumption might be incorrect for some columns,
%GenHap does not require the \textit{all-heterozygous} assumption, differently from ReFHap \cite{duitama2010} and ProbHap \cite{kuleshov2014ProbHap}.
%So, we allow also for the presence of a small number of homozygous sizes in the solution.
and these algorithms can only deal with limited reads coverages.
For example, 
%on the one hand, 
ProbHap \cite{kuleshov2014ProbHap} can handle long reads coverage values up to $20\times$, which is not appropriate for higher coverage short-read datasets;
on the other hand, it works better with very long reads at a relatively shallow coverage ($\leq 12\times$).

More recently, 
%relying on the effectiveness achieved by dynamic programming algorithms \cite{he2010}, WhatsHap \cite{patterson2015} was proposed.
a tool based on a dynamic programming approach, called WhatsHap, was presented \cite{patterson2015}.
WhatsHap is based on a fixed parameter tractable algorithm \cite{he2010, bonizzoni2015}, and leverages the long-range information of long reads; however,
%since it is linear in the read length.
 it can deal only with datasets of limited coverage up to $\sim 20 \times$.
A parallel version of WhatsHap has been recently proposed in \cite{bracciali2016}, showing the capability to deal with higher coverages up to $\sim 25 \times$.
An alternative approach, called HapCol \cite{pirola2015}, uses the uniform distribution of sequencing errors characterizing long reads.
%Especially, the authors propose
In particular, HapCol exploits a new formulation of the wMEC problem, where the maximum number of corrections is bounded in every column and is computed from the expected error rate.
%Moreover, 
HapCol can only deal with instances of relatively small coverages up to $\sim 25-30 \times$.
    
To sum up, even though high-throughput DNA sequencing technologies are paving the way for valuable advances in clinical practice, analyzing such an amount of data still represents a challenging task.
This applies especially to clinical settings, where accuracy and time constraints are critical \cite{rimmer2014}. 

In order to tackle the computational complexity of the haplotyping problem, in this work we propose GenHap, a novel computational method for haplotype assembly based on Genetic Algorithms (GAs).
GenHap can efficiently solve large instances of the wMEC problem, yielding optimal solutions by means of a global search process, without any \textit{a priori} hypothesis about the sequencing error distribution in reads.
The computational complexity of the problem is overcome by relying on a \textit{divide-et-impera} approach, which provides faster and more accurate solutions compared with the state-of-the-art haplotyping tools.
The paper is structured as follows. In the next section, we briefly introduce the haplotyping problem, and describe in detail the GenHap methodology along with its implementation.
Then, we show the computational performance of GenHap, extensively comparing it against HapCol.
We finally provide some conclusive remarks and future improvements of this work.

\section*{Methods}

\subsection*{Problem formulation}

Given $n$ positions on two homologous sequences belonging to a diploid organism and $m$ reads obtained after a sequencing experiment, we can reduce each read to a fragment vector $\mathbf{f} \in \{0,1,-\}^n$, where $0$ denotes a position that is equal to the reference sequence, $1$ denotes a SNP with respect to the reference sequence and $-$ indicates a position that is not covered by the read.
We define a haplotype as a vector $\mathbf{h} \in \{0,1\}^n$, that is, the combination of SNPs and wild-type positions belonging to one of the two chromosomes.
% \todo[inline]{Ennesima domanda cretina, scusate. In che senso wild-type? Capisco cosa intendiamo, ma si dice davvero cosi? (MS)
%Yees, sono definiti wild-type le posizioni uguali al genoma di riferimento (A)}
Given the two haplotypes $\mathbf{h}_{1}$ and $\mathbf{h}_{2}$---which refer to the first and second copy of the chromosome, respectively---a position $j$ (with $j \in \{1, \dots, n\}$) is said to be heterozygous if and only if $h_{1_j} \neq h_{2_j}$, otherwise $j$ is homozygous.
% \todo[inline]{"THE" two haplotypes, nel senso che si riferiscono ai due cromosomi? Specificare meglio, please (D)
% Ci sono due copie dello stesso cromosoma, quindi ci sono due aplotipi (A)}
% Let the ``fragment matrix'' be the $m \times n$ matrix $\mathbf{M}$ containing all the fragments.
% \textcolor{red}{Note that $\mathbf{h}_{1}$ and $\mathbf{h}_{2}$ refer to the first and second copy of the chromosome, respectively.}

\begin{figure}[t!]
  \includegraphics[width=0.97\textwidth]{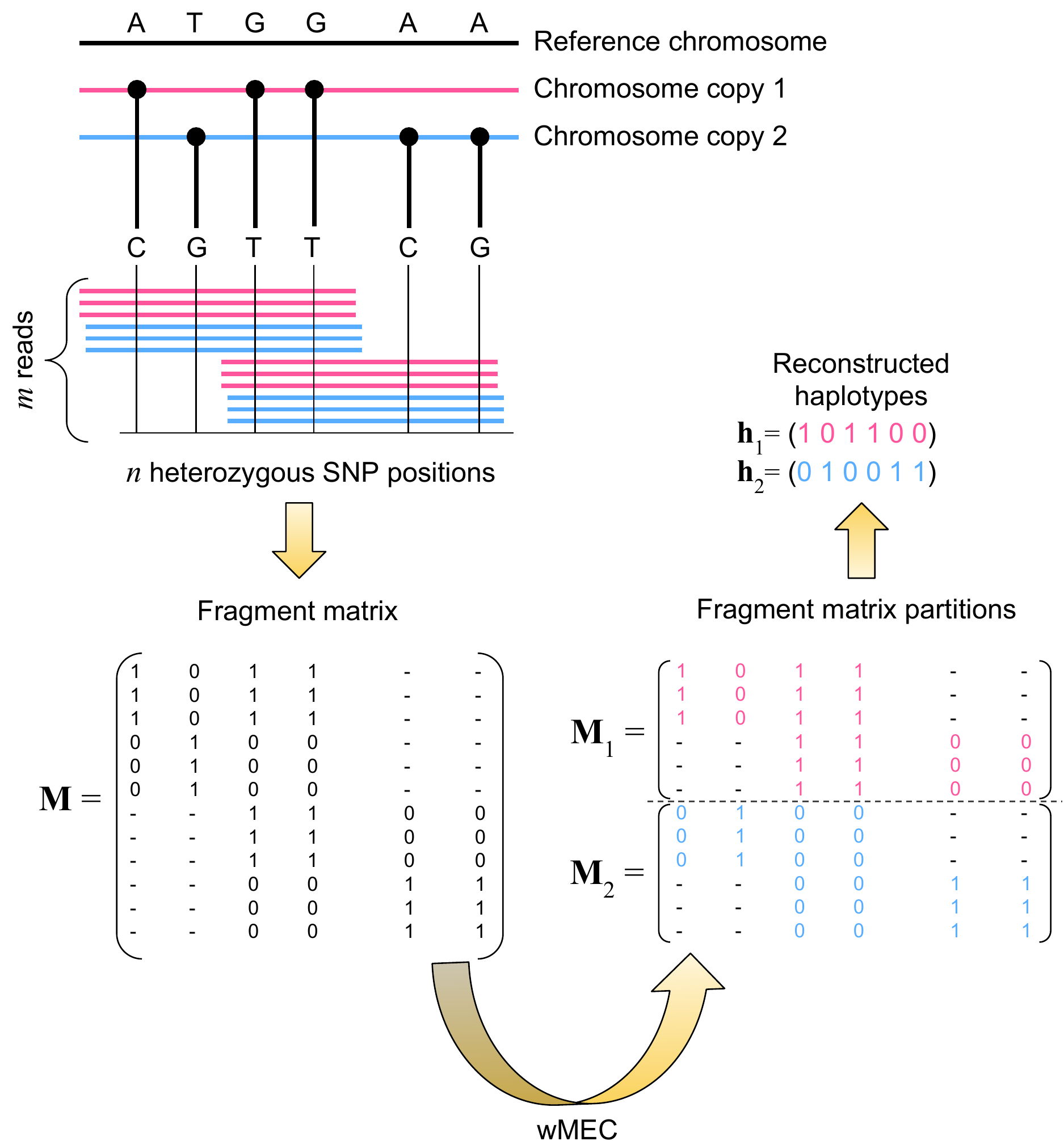}
  \caption{Simplified workflow of the haplotype assembly process. Raw sequencing data are initially aligned, defining $m$ reads.
Every position of the two chromosome copies is compared against a reference chromosome. The black solid points denote $n$ heterozygous positions, along with the corresponding nucleobases.
The fragment matrix $\mathbf{M}$ is defined assigning $1$ to SNP positions and $0$ to wild-type positions.
To reconstruct the two haplotypes $\mathbf{h}_1$ and $\mathbf{h}_2$ characterized by the least number of corrections to the SNP values among the $2^n$ candidate haplotypes, the wMEC problem is solved by partitioning the matrix $\mathbf{M}$ into two disjoint matrices $\mathbf{M}_1$ and $\mathbf{M}_2$.
% \todo[inline]{Domanda stupida, ma per capire. Nella figura il problema non e' gia' risolto all'inizio del processo? le read sono già separate in colori a assegnate alle rispettive copie dei cromosomi. Anche la fragment matrix e' gia separata correttamente, non capisco cosa succeda nel box haplotype assembly a queste strutture dati. Immagino anche ci sia una ratio per il fatto che i nucleotidi siano distanziati irregolarmente (MS)
% I nucleotidi sono distanziati irregolarmente per ripecchiare la realtà, le read sono colorate per aiutare il lettore.
% Nella matrice ci sono i colori sempre per aiutare il lettore (A)}
%\todo[inline]{"we can reconstruct ... with the least number of corrections to the SNP values": we, chi? Noi con GenHap, o in generale? Se è con GenHap, allora attenzione perche' questa figura e' citata nel testo nel punto in cui si sta descrivendo il problema di haplotyping, NON dove si descrive come funziona GenHap (D) è il processo generale del haplotyping assembly, quindi vale per tutti i metodi che lavorano direttamente con le reads (A)}
}
  \label{fig:ReadsMatHap}
\end{figure}

Let $\mathbf{M}$ be the ``fragment matrix'', that is, the $m \times n$ matrix  containing all fragments.
Two distinct fragments $\mathbf{f}$ and $\mathbf{g}$ are said to be in conflict if there is a position $j$ (with $j \in \{1, \dots, n\}$) such that $f_{j} \neq g_{j}$ and $f_{j}, g_{j} \neq -$, otherwise they are in agreement.
$\mathbf{M}$ is conflict-free if there are two different haplotypes $\mathbf{h}_1$ and $\mathbf{h}_2$ such that each row $M_i$ (with $i \in \{1, \dots, m\}$) is in agreement with either $\mathbf{h}_1$ or $\mathbf{h}_2$.
The overall haplotype assembly process is outlined in Figure \ref{fig:ReadsMatHap}.

We can extend the heterozygous and homozygous definition at the column level as follows: a column $c$ of $\mathbf{M}$ is homozygous if all its values are either in $\{0,-\}$ or in $\{1,-\}$, on the contrary $c$ is heterozygous because its values are in $\{0,1,-\}$ meaning that both a SNP and a wild-type exist in that position.
%\todo[inline]{both a SNP and a wild-type oppure AT LEAST a SNP and a wild-type? (D)
%è giusto a SNP and a wild-type, perché uno deve avere il valore 0 (wild-type) e l'altro 1 (SNP).
%Essendo diploidi gli unici due valori sono quelli (A)}
Finally, we can detect the case where two distinct fragments are in conflict, and measure their diversity by defining a distance $D(\cdot,\cdot)$ that calculates the number of different values between two fragments.
Namely, given $\mathbf{f}=(M_{i1}, \dots, M_{in})$ and $\mathbf{g}=(M_{l1}, \dots, M_{ln})$ of $\mathbf{M}$ (with $i,l \in \{1, \dots, m\}$), we consider:
	\begin{equation}
    \label{eq:distance}
	D(\mathbf{f}, \mathbf{g}) = \sum_{j=1}^{n} d(f_j, g_j),
	\end{equation}
	where $d(f_j, g_j)$ is defined as:
	\begin{equation}
    \label{eq:HamDist}
	d(x,y) = \begin{cases} 1, & \mbox{if } x \neq y, \mbox{ } x \neq -, \mbox{ and } y \neq -\\ 0, & \mbox{otherwise} \end{cases}.
	\end{equation}
%	If $\mathbf{M}$ is conflict-free and there are no errors in all reads, $\mathbf{M}$ can be divided into two disjoint partitions $\mathbf{M}_1$ and $\mathbf{M}_2$ containing the fragments that are conflict-free, thus allowing to infer the two haplotypes, the first one from $\mathbf{M}_1$ and the second one from $\mathbf{M}_2$.

Equation~(\ref{eq:distance}) defines the \textit{extended Hamming distance} between two ternary strings $\mathbf{f}$ and $\mathbf{g}$ \cite{chen2013}, denoting the total number of positions wherein both characters of $\mathbf{f}$ and $\mathbf{g}$ belong to $\{0,1\}$ but they are different according to Equation~(\ref{eq:HamDist}).

If $\mathbf{M}$ is conflict-free, then it can be partitioned into two disjoint matrices $\mathbf{M}_1$ and $\mathbf{M}_2$, each one containing a set of conflict-free fragments.
We can infer the two haplotypes $\mathbf{h}_1$ and $\mathbf{h}_2$ from $\mathbf{M}_1$ and $\mathbf{M}_2$, respectively, as follows:	 
%	Let $\mathbf{h}_1$ and $\mathbf{h}_2$ be two possible haplotypes, $\mathbf{M}_1$ and $\mathbf{M}_2$ the partitions of $\mathbf{M}$ representing the classification of all SNP fragments.
%	We can infer $\mathbf{h}_1$ and $\mathbf{h}_2$ from $\mathbf{M}_1$ and $\mathbf{M}_2$, respectively, as follows:
\begin{equation}
	\label{eq:haploCal}
	h_{k_j} = \begin{cases} 1, & \mbox{if } N_{1_j}(\mathbf{M}_k) \geq N_{0_j}(\mathbf{M}_k)\\ 0, & \mbox{otherwise} \end{cases},
\end{equation}
where $j \in \{1, \dots, n\}$, $k\in \{1,2 \}$, and $N_{0_j}(\mathbf{M}_k)$, $N_{1_j}(\mathbf{M}_k)$ denote the number of $0$s and $1$s in the $j$-th column, respectively.
In such a way, $\mathbf{N}_0(\mathbf{M}_k)$ is the vector consisting of the number of $0$s of each column $j$ using the reads of the partition $\mathbf{M}_k$, while $\mathbf{N}_1(\mathbf{M}_k)$ is the vector consisting of the number of $1$s of each column $j$ represented by the partition $\mathbf{M}_k$.

In order to solve the wMEC problem, $\mathbf{N}_0$ and $\mathbf{N}_1$ are calculated using the $m \times n$ weight matrix $\mathbf{W}$, representing the weight associated with each position in each fragment.
As a matter of fact, $\mathbf{W}$ can be divided into the two disjoint partitions $\mathbf{W}_1$ and $\mathbf{W}_2$, whose row indices correspond to those in $\mathbf{M}_1$ and $\mathbf{M}_2$, respectively.
We can extend Equation~(\ref{eq:haploCal}) taking into account the weights as follows:
\begin{equation}
	\label{eq:haploCalWeigh}
	h_{k_j} = \begin{cases} 1, & \mbox{if } N_{1_j}(\mathbf{W}_k) \geq N_{0_j}(\mathbf{W}_k)\\ 0, & \mbox{otherwise} \end{cases},
\end{equation}
where $j \in \{1, \dots, n\}$, $k\in \{1,2 \}$, and $N_{0_j}(\mathbf{W}_k)$, $N_{1_j}(\mathbf{W}_k)$ denote the sum of the weights associated with the $0$ and $1$ elements in the $j$-th column, respectively.

The distance $D(\cdot, \cdot)$ given in Equation~(\ref{eq:distance}) can be used also to evaluate the distance between a fragment and a haplotype, by means of the following error function:
\begin{equation}
	\label{eq:fitness}
	\mathcal{E}(\mathbf{M}_1,\mathbf{M}_2, \mathbf{h}_1, \mathbf{h}_2) = \sum_{k=1}^{2} \sum_{\mathbf{f} \in \mathbf{M}_k} D(\mathbf{f}, \mathbf{h}_k).
\end{equation}
The best partitioning of $\mathbf{M}$ can be obtained by minimizing Equation (\ref{eq:fitness}), inferring $\mathbf{h}_1$ and $\mathbf{h}_2$ with the least number of errors.
Equation~(\ref{eq:fitness}) is used as fitness function in GenHap.

\subsection*{GenHap: haplotype assembly using GAs}
GAs are population-based optimization strategies mimicking Darwinian processes \cite{goldberg1989,baker1985,miller1995}.
In GAs, a population $P$ of randomly generated individuals undergoes a selection mechanism and is iteratively modified by means of genetic operators (i.e., crossover and mutation).
Among the existing meta-heuristics for global optimization, GAs are the most suitable technique in this context thanks to the discrete structure of the candidate solutions.
This structure is well-suited to efficiently solve the intrinsic combinatorial nature of the haplotype assembly problem.
% \todo[inline]{da "and" in avanti mi chiedo: EH? (P) Non ho capito cosa non si capisce. (A)}
In the most common formulation of GAs, each individual $C_p$ (with $p \in \{1, \ldots, |P|\}$) encodes a possible solution of the optimization problem as a fixed-length string of characters taken from a finite alphabet.
Based on a quality measure (i.e., the fitness value), each individual is involved in a selection process in which individuals characterized by good fitness values have a higher probability to be selected for the next iteration.
Finally, the selected individuals undergo crossover and mutation operators to possibly improve offspring and to introduce new genetic material in the population.
%Starting from the selected chromosomes, new ones are generated, named offspring, by applying genetic operations (i.e., crossover and mutation), which allow to mix the parents' characteristics introducing new genetic material for enhancing exploration capabilities.
% A general scheme for GAs can be summarized as follows:
% \begin{enumerate}
%     \item a population of individuals representing solutions of a given problem is randomly generated;
% 	\item the fitness function of each individual is evaluated;
% 	\item the individuals are selected according to their fitness value;
% 	\item crossover and mutation operators are applied with crossover rate $c_r$ and mutation rate $m_r$, respectively, to recombine and vary the selected individuals;
% 	\item if a chosen termination criterion is not satisfied, go back to step 2; otherwise, the individual characterized by the best fitness value is returned.
% \end{enumerate}
% \todo[inline]{secondo me questo general scheme si può togliere (P)
% Se vogliamo asciugare la parte sul GA si può togliere (A)}

GenHap exploits a very simple and efficient structure for individuals, which encodes as a binary string a partition of the fragment matrix $\mathbf{M}$.
%The length of each individual $C_p$, $p = 1, 2, \dots, |P|$ is equal to the number of reads $m$. %, and each of its elements is a binary value referring to a read.
In particular, each individual $C_p=[C_{p_1}, C_{p_2}, \ldots, C_{p_m}]$ (with $p \in \{1, \ldots, |P|\}$) is encoded as a circular array of size $m$ (i.e., the number of reads).
In order to obtain the two partitions $\mathbf{M}_1$ and $\mathbf{M}_2$, $C_p$ is evaluated as follows: if the $i$-th bit is equal to $0$, then the read $i$ belongs to $\mathbf{M}_1$; otherwise, the read $i$ belongs to $\mathbf{M}_2$.
Once the two partitions are computed, GenHap infers the haplotypes $\mathbf{h}_1$ and $\mathbf{h}_2$ by applying Equation~(\ref{eq:haploCalWeigh}).
Finally, Equation~(\ref{eq:fitness}) is exploited to calculate the number of errors made by partitioning $\mathbf{M}$ as encoded by each individual of $P$.
This procedure is iterated until the maximum number of iterations $T$ is reached, the number of errors is equal to $0$ or the fitness value of the best individual does not improve for $\theta = \lceil 0.25 \cdot T \rceil$ iterations.
%In our approach, we used a population composed of $100$ individuals. 
	
Among the different selection mechanisms employed by GAs (e.g., roulette wheel \cite{goldberg1989}, ranking \cite{baker1985}, tournament \cite{miller1995}), GenHap exploits the tournament selection to create an intermediate population $P'$, starting from $P$.
In each tournament, $\kappa$ individuals are randomly selected from $P$ and the individual characterized by the best fitness value is added to $P'$.
The size of the tournament $\kappa$ is related to the selection pressure: if $\kappa$ is large, then the individuals characterized by worse fitness values have a low probability to be selected, therefore the variability of $P'$ might decrease. %; moreover, according to their fitness values, multiple copies of the same individuals could be added to $P'$.

Afterwards, the genetic operators (i.e., crossover and mutation) are applied to the individuals belonging to $P'$ to obtain the offspring for the next iteration.
GenHap exploits a single-point crossover with mixing ratio equal to $0.5$. 
Crossover is applied with a given probability $c_r$ and allows for the recombination of two parent individuals $C_{y}, C_{z} \in P'$ (for some $y, z \in \{1, \ldots, |P|\}$), generating two offspring that possibly have better characteristics with respect to their parents.
%This operator is applied with a given probability $c_r$ to the individuals selected by the tournament strategy and belonging to the first intermediate population $P'$.
%In order to blend exactly half characters from $m_r$ and half from $C_f$, we use a single-point crossover resulting in a mixing ratio fixed to $0.5$.
%Namely, given two parent individuals $C_{y}, C_{z} \in P'$ (for some $y, z = 1, \ldots, |P|$), a crossover point $\chi$ is randomly selected from the circular arrays $[C_{y_1}, C_{y_2}, \ldots, C_{y_m}]$ and $[C_{z_1},C_{z_1}, \ldots, C_{z_m}]$ encoding the individuals, and two offspring are generated by exchanging $50\%$ of the elements between the two parents.
%The offspring are then inserted into a second intermediate population $P''$.

In order to increase the variability of the individuals, one or more elements of the offspring can be modified by applying the mutation operator.
GenHap makes use of a classic mutation in which 
%a random number $rnd$ is generated in $[0,1)$ for each 
the elements $C_{p_e}$ (with  $e \in \{1, \dots, m\}$) of the individual can be flipped  (i.e., from $0$ to $1$ or vice-versa) with probability $m_r$.
%of the selected individual belonging to $P''$.
%If $rnd$ is lower than $m_r$, the element $C_{p_e}$ is mutated by flipping its value (i.e., from $0$ to $1$ or vice-versa).
Besides this mutation operator, GenHap implements an additional bit-flipping mutation in which a random number of consecutive elements of the individual is mutated according to probability $m_r$. 
This operator is applied if the fitness value of the best individual does not improve for a given number of iterations ($2$ in our tests).
 
 \begin{figure}[!ht]
  \includegraphics[width=0.95\textwidth]{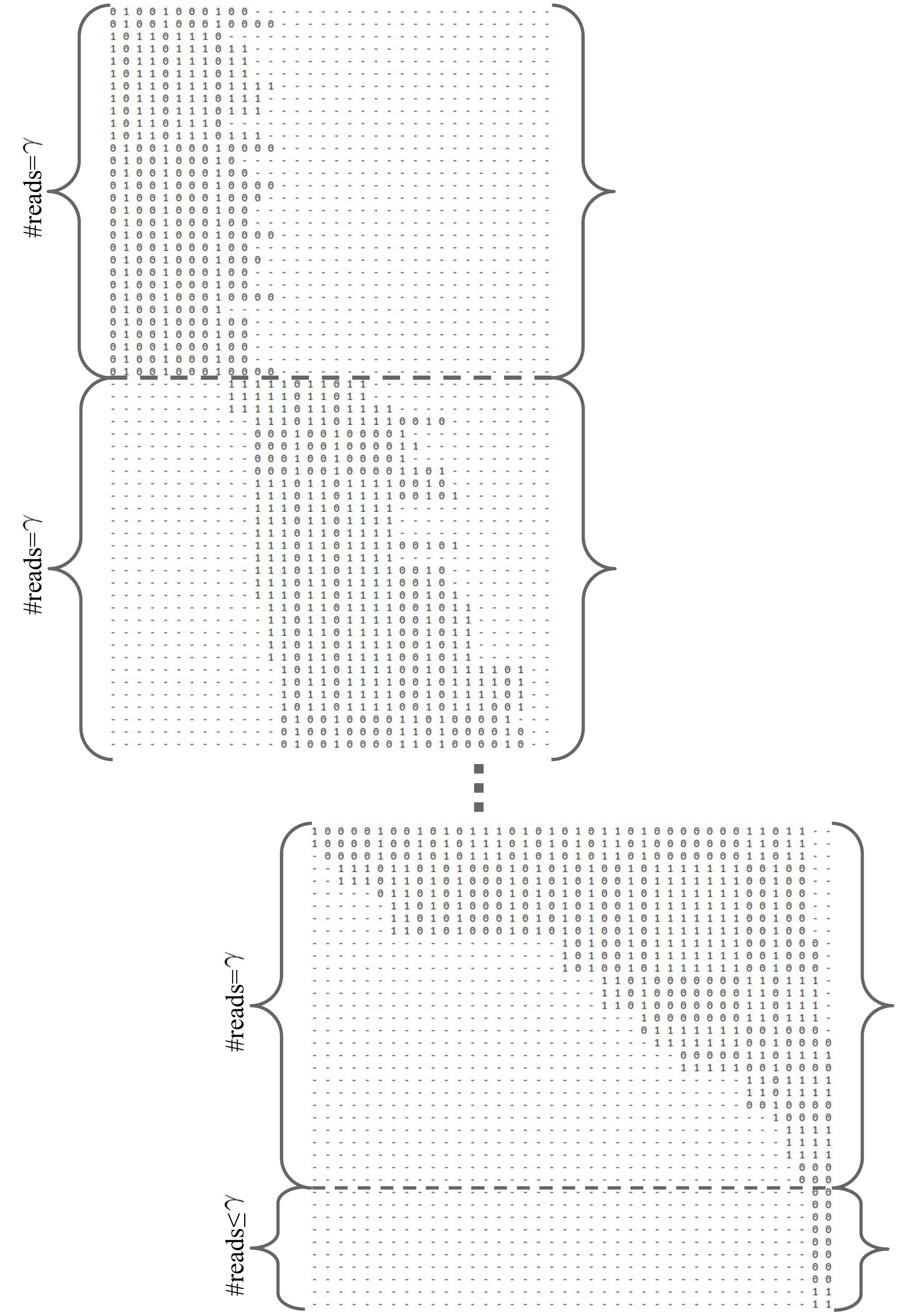}
  \caption{Scheme of the partition of the input matrix: the input matrix $\mathbf{M} \in \{0,1,-\}^{m \times n}$ is split into sub-matrices consisting of $\gamma$ reads, generating $\Pi = \lfloor m/\gamma \rfloor$ sub-problems that are solved independently by a GA instance.
The last sub-matrix could have a number of reads lower than $\gamma$.}
  \label{fig:matrixDivision}
\end{figure}
 
Finally, to prevent the quality of the best solution from decreasing during the optimization, GenHap exploits an elitism strategy, so that the best individual from the current population is copied into the next population without undergoing the genetic operators.
    
Unlike the work in \cite{wang2005}, GenHap solves the wMEC problem instead of the unweighted MEC formulation, by means of Equation~(\ref{eq:haploCalWeigh}).
Moreover, differently from the other heuristic strategies, such as ReFHap \cite{duitama2010} and ProbHap \cite{kuleshov2014ProbHap}, we did not assume the all-heterozygosity of the phased positions \cite{chen2013}.
Under this assumption, every column corresponds to heterozygous sites, implying that $\mathbf{h}_1$ must be the complement of $\mathbf{h}_2$.
% \todo[inline]{Questa dovrebbe essere una spiegazione della novelty di GenHap, giusto? Se è così non è chiarissimo. Secondo me andrebbe spiegato meglio (P)}
In addition, since the required execution time as well as the problem difficulty increase with the number of reads and SNPs, to efficiently solve the wMEC problem we split the fragment matrix $\mathbf{M}$ into $\Pi = \lfloor m/\gamma \rfloor$ sub-matrices consisting of $\gamma$ reads (see Figure~\ref{fig:matrixDivision}).
Following a  \textit{divide-et-impera} approach \cite{maisto2015}, the computational complexity can be tackled by partitioning the entire problem into smaller and manageable sub-problems, each one solved by a GA that converges to a solution characterized by two sub-haplotypes with the least number of corrections to the SNP values. 
The solutions to the sub-problems achieved by the $\Pi$ GA instances are finally combined.
%whose solutions are eventually combined \cite{maisto2015}. 
%As a matter of fact, by considering these sub-problems, each GA execution can improve the convergence to a solution characterized by two sub-haplotypes with the least number of corrections to the SNP values.
This approach is feasible thanks to the long reads with higher coverage produced by the second- and third-generation sequencing technologies.
% \todo[inline]{questo non cozza con il fatto che poi consideriamo r454 che, come scritto sotto, produces short but precise reads? (P)}
As a matter of fact, highly overlapping reads allow us to partition the problem into easier sub-problems, avoiding the possibility of obtaining incorrect reconstructions during the merging phase.
%In these sub-problems, each different GA instance has a fast convergence for reconstructing the two sub-haplotypes with the least number of corrections to the SNP values.
%So doing, the computational hardness of the problem can be tackled relying on a \textit{divide-et-impera} approach.
%The whole complex problem is partitioned into smaller and more manageable sub-goals, which are then combined \cite{maisto2015}.

The parameter $\gamma$, used for the calculation of $\Pi$, depends on the coverage value and on the nature of the sequencing technology; its value must be set to avoid discrete haplotype blocks that do not exist in the input matrix $\mathbf{M}$.
Generally, the intervals where several independent historical recombination events occurred separate discrete blocks, revealing greater haplotype diversity for the regions spanning the blocks \cite{daly2001}.
% The haplotype blocks generally reveal haplotype diversity for regions spanning the blocks \cite{daly2001}.
% \todo[inline]{sì ma la ragione non è scritta (P)}

GenHap firstly detects all the haplotype blocks inside the fragment matrix $\mathbf{M}$ and then, in each block, it automatically sets $\gamma$ equal to the mean coverage of that block to partition the reads.
Notice that GenHap solves each block sequentially and independently, obtaining a number of haplotype pairs equal to the number of detected blocks.
So doing, for each block GenHap proceeds by executing $\Pi$ different GA optimizations, one for each sub-problem, calculating $2\cdot\Pi$ sub-haplotypes.
The length of the individuals is equal to $\gamma$, except for the last sub-problem that could have a number of reads smaller than $\gamma$ (accordingly, the length of the individuals could be smaller than $\gamma$).

Since the problem is divided into $\Pi$ sub-problems, two sub-problems referring to contiguous parts of the two chromosome copies might contain some overlapped positions that can be either homozygous or heterozygous.
However, the reads covering an overlapped position might not be entirely included in the same sub-problem.
For this reason, during the GA-based optimizations, all the phased positions are assumed to be heterozygous.
If a position $j$ is homozygous (i.e., all the reads covering this position have the same value, belonging to $\{0, -\}$ or $\{1, -\}$, in both the sub-partitions and in every read covering it), then only one of the two sub-haplotypes will have the correct value.
This specific value is correctly assigned to the sub-haplotype covered by the highest number of reads by following a majority rule.
As soon as the two sub-haplotypes are obtained, all the possible uncorrected heterozygous sites are removed and the correct homozygous values are assigned by checking the columns of the two sub-partitions.
% Since the problem is divided in $\Pi$ sub-problems, two sub-problems referring to contiguous parts of the chromosome might contain some overlapped positions that can be either homozygous or heterozygous, but not all the reads covering the same position will be in the same sub-problem.
% For this reason, during the optimizations all phased positions are assumed to be heterozygous.
% If a position $j$ is homozygous (i.e., all the reads covering this position have the same value) only one of the two sub-haplotypes will have the correct value, which is assigned exploiting a major rule that sets the correct value in the sub-haplotype covered by the highest number of reads.
% As soon as the two sub-haplotyes are obtained, all the possible uncorrected heterozygous sites are removed and the correct homozygous values are assigned by checking the columns of the two sub-partitions.
% We highlight that a position $j$ is homozygous if it assumes the same value ($0$ or $1$) in both the sub-partitions and in all the reads covering it.
Finally, once all sub-problems in $\Pi$ are solved, GenHap recombines the sub-haplotypes to obtain the two entire haplotypes $\mathbf{h}_1$ and $\mathbf{h}_2$ of the block under analysis.
%To be more precise, GenHap merges the sub-haplotypes calculated in the first sub-problem with those calculated in the second problem, and so on. 

GenHap is also able to find and mask the ambiguous positions by replacing the $0$ or $1$ value with a $X$ symbol.
We highlight that an ambiguous position is a position covered only by the reads belonging to one of the two haplotypes.
% \todo[inline]{non si capisce cos'è una posizione ambigua e a cosa serve mettere la x (P)}

\subsection*{Implementation}

% \begin{figure}[!t]
%   \includegraphics[width=0.95\textwidth]{images/GenHapMasterSlaveScheme.pdf}
%   \caption{Scheme of the Master-Slave implementation of GenHap: the Master process orchestrates all the $\Sigma$ Slaves sending one or more sub-partitions to each Slave, which then solves the assigned wMEC sub-task.}
%   \label{fig:MS_GenHap}
% \end{figure}

\begin{figure}[ht]
  \includegraphics[width=0.95\textwidth]{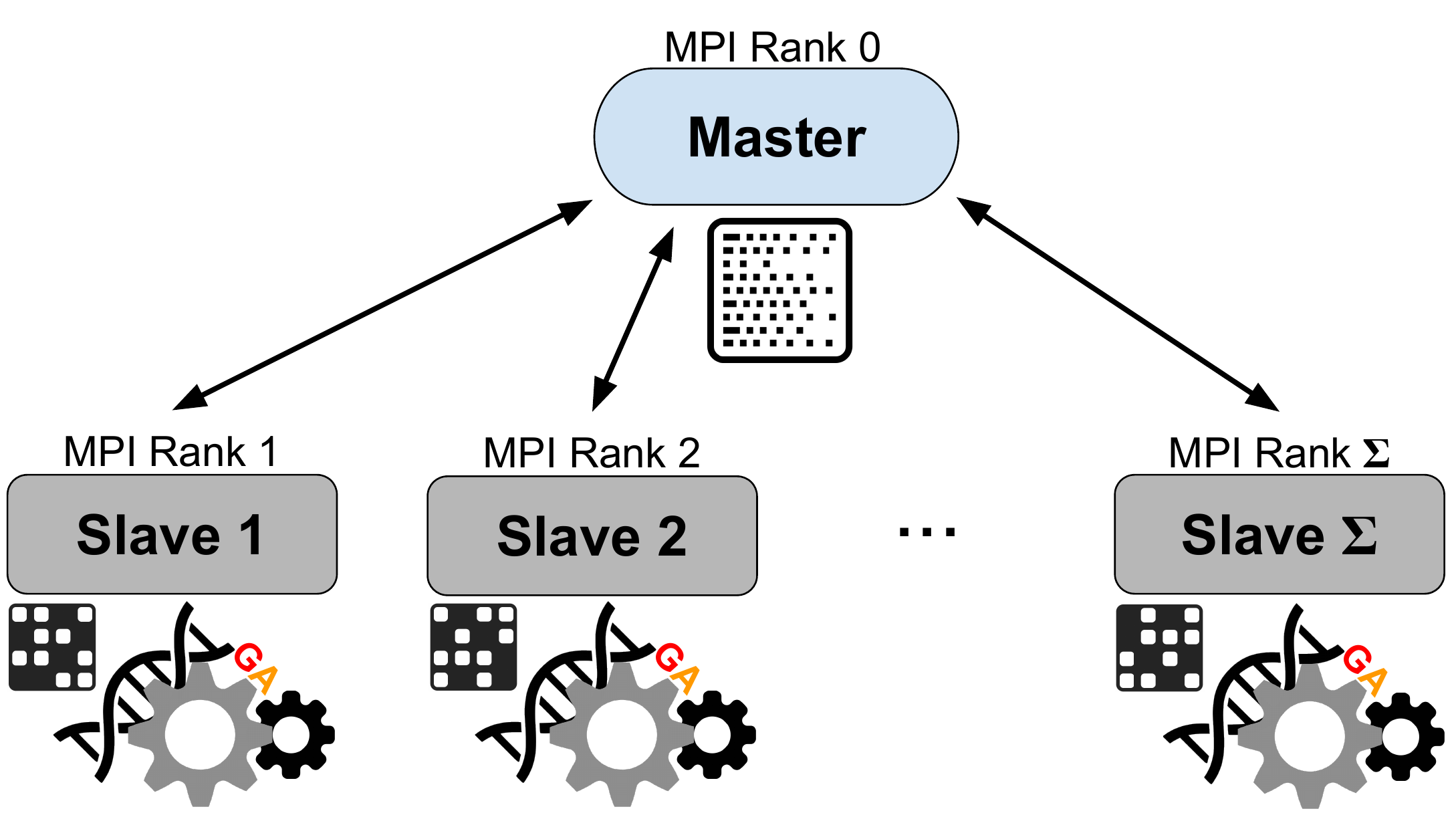}
  \caption{Scheme of the Master-Slave implementation of GenHap: the Master process orchestrates all the $\Sigma$ Slaves sending one or more sub-partitions to each Slave, which then solves the assigned wMEC sub-task.}
  \label{fig:MS_GenHap}
\end{figure}

In order to efficiently solve the wMEC problem and tackle its computational complexity, 
GenHap detects the haplotype blocks inside the matrix $\mathbf{M}$ and then, for each block, it splits the portion of $\mathbf{M}$ into $\Pi$ sub-matrices consisting of $\gamma$ reads.
So doing, the convergence speed of the GA is increased thanks to the lower number of reads to partition in each sub-problem with respect to the total number of reads of the whole problem.
% \todo[inline]{Come sopra, anche qui non si capisce perché dovrebbe aumentare la velocità di convergenza (P)}
As shown in Figure~\ref{fig:MS_GenHap}, the $\Pi$ sub-matrices are processed in parallel by means of a \textit{divide-et-impera} approach that exploits a Master-Slave distributed programming paradigm \cite{tangherloni2018PDP} to speed up the overall execution of GenHap.
This strategy allowed us to distribute the computation in presence of multiple cores.
As a matter of fact, GenHap works by partitioning the initial set of reads into sub-sets and solving them by executing different GA instances.
This strategy can be exploited in GenHap, as it solves the wMEC problem working on the rows of the fragment matrix $\mathbf{M}$; on the contrary, HapCol works considering the columns of $\mathbf{M}$, which cannot be independently processed in parallel.

The functioning of our Master-Slave implementation can be summarized as follows:
\begin{enumerate}
 \item the Master allocates the resources and detects the haplotype blocks inside the fragment matrix.
For each detected block it partitions the portion of the matrix $\mathbf{M}$ into $\Pi$ sub-matrices and offloads the data onto the available $\Sigma$ Slaves (in real scenarios, $\Sigma \ll \Pi$).
During this phase, each Slave generates the initial population of the GA;
 \item the $\sigma$-th Slave (with $\sigma \in \{1, \dots, \Sigma\}$) executes the assigned wMEC sub-task, running the GA for either $\theta$ non-improving iterations or $T$ maximum iterations, independently of the other Slaves;
 \item the process is iterated until all the wMEC sub-tasks are terminated;
 \item the Master recombines the sub-solutions received from the Slaves, and returns the complete wMEC solution for the block under analysis.
\end{enumerate}

GenHap was entirely developed using the C++ programming language exploiting the Message Passing Interface (MPI) specifications to leverage multi-core Central Processing Units (CPUs).

\section*{Results}

In this section we first describe the synthetic and real datasets used during the tests and present the results obtained to identify the best GA setting.
Then, we discuss the performance achieved by GenHap with respect to HapCol \cite{pirola2015}, which was previously shown to be more efficient than the other existing methods for the haplotype assembly problem, both in terms of memory consumption and execution time.
%\todo[inline]{Specificare che abbiamo anche usato dati reali e descrivere le caratteristiche di Genome-In-A-Bottle (GIAB) \cite{zook2014} nella sottosezione successiva.\\
%Descrivere la procedura del calcolo dell'agreement con i risultati di HapCol. (Leo)}

\subsection*{The analyzed datasets}
%GenHap is designed for third-generation sequencing technologies, which produce longer reads with higher coverage.
%Unfortunately, such a kind of data is not still widely available.
In order to test the performance of GenHap, we generated two synthetic (yet realistic) datasets, each one consisting of instances obtained from a specific sequencing technology.
In particular, we considered the Roche/454 genome sequencer (Roche AG, Basel, Switzerland), representing one of the next-generation sequencing (NGS) systems able to produce long and precise reads, and the PacBio RS II sequencer \cite{roberts2013, carneiro2012}, which is an emerging third-generation sequencing technology.
Note that the reads produced by the Roche/454 sequencer are approximately $9$-times shorter than those generated by the PacBio RS II system.

In order to generate the datasets, we exploited the General Error-Model based SIMulator (GemSIM) toolbox \cite{mcelroy2012}.
GemSIM is a software able to generate \textit{in silico} realistic sequencing data.
It relies on empirical error models and distributions learned from real NGS data, and simulates both single‐ and paired‐end reads from a single genome, collection of genomes, or set of related haplotypes.
GemSIM can in principle simulate data from any sequencing technology producing output data encoded in the FASTQ format \cite{cock2009}, for raw reads, and Sequence Alignment/Map (SAM), for aligned reads. %, including, for instance, Illumina and Roche/454 sequencing technologies.
In this work, we exploited the error model for the Roche/454 sequencer, already available in GemSIM, and defined an additional error model for the PacBio RS II technology.
The synthetic reads were generated from the reference sequence of the human chromosome 22 (UCSC Genome Browser, GRCh37/hg19 Feb. 2009 assembly \cite{casper2017ucsc}), in which random SNP were inserted.

%In addition, we defined a new error model for a the third-generation sequencing methods, namely the PacBio system.
We exploited the GemHaps tool included in GemSIM \cite{{mcelroy2012}} to generate a haplotype file starting from a given genome sequence, and specifying the number as well as the frequency of SNPs in each haplotype, denoted by $\#\text{SNPs}$ and $f_\text{SNPs}$, respectively.
%We exploited the GemHaps that, given a genome sequence and the specifications on haplotype frequencies and the number of SNPs in each haplotype (defined as a group of related nucleotide sequences, each one differing by at least one SNP), yields a haplotype file.
Note that the SNP positions were randomly determined.
% Then, the resulting haplotype file was processed by GemReads, together with an error model file (generated by GemErr or supplied in GemSIM), a FASTA genome file (or directory), a tab-delimited species-abundance text file (metagenomics mode only), and the selected quality score offset.
Then, the resulting haplotype file was processed by GemReads, together with an error model file (generated by GemErr or supplied in GemSIM), a FASTA genome file (or directory), and the selected quality score offset.
The resulting SAM file was converted into the compressed Binary Alignment/Map (BAM) format for a more efficient manipulation \cite{li2009}.
%The BAM file conveys the same information of SAM format in a compressed format.
In order to store the SNPs, we exploited the Variant Call Format (VCF) \cite{danecek2011}, which is the most used format that combines DNA polymorphism data, insertions and deletions, as well as structural variants.
Lastly, the BAM and VCF files were processed to produce a WhatsHap Input Format (WIF) file \cite{patterson2015}, which is the input of GenHap.
%\todo[inline]{io tutta sta parte la farei molto più corta (S)}

The two synthetic datasets are characterized by the following features: \textit{i}) $\#\text{SNPs} \in \{500, 1000, 5000, 10000, 20000\}$ (equally distributed over the two haplotypes); \textit{ii}) coverage $\text{cov} \in \{\sim\! 30\times$,  $\sim\!60\times\}$; \textit{iii}) average $f_\text{SNPs} \in \{100, 200\}$, which means one SNP every $100$bp or $200$bp \cite{nachman2001single,gabriel2002structure}, varying the portion of genome onto which the reads were generated.
Read lengths were set to $600$bp and $5000$bp for the Roche/454 and the PacBio RS II sequencers, respectively. %, while we consider a SNP frequency of one SNP every $100$bp or $200$bp.
The number of reads was automatically calculated according to the value of $\text{cov}$ and the sequencing technologies, by means of the following relationship:
\begin{equation}
\label{eq:reads}
\#\text{reads} = \text{cov} \cdot\frac{len(\text{genome})}{len(\text{read})},
\end{equation}
where $len(\text{genome})$ represents the length of the considered genome, which starts at a given position $x$ and ends at position $y = x + f_\text{SNPs}\cdot\#\text{SNPs}$.

In order to test the performance of GenHap on real sequencing data, we exploited a WIF input file present in \cite{Beretta170225}, which was generated starting from high-quality SNP calls and sequencing data made publicly available by the Genome in a Bottle (GIAB) Consortium \cite{zook2014}.
In particular, we exploited data produced by the PacBio technology and limited to the chromosome $22$ of the individual NA12878.
real dataset available at
Moreover, we tested GenHap on an additional  \cite{pacbio54}.
As for the previous dataset, we limited our analysis to chromosome $22$.
The available BAM file--containing long reads with high-coverage produced with the PacBio RS II sequencing technology--and the VCF file were processed to obtain a WIF input file as described above.

\subsection*{GA setting analysis}

\begin{figure}[t!]
  \includegraphics[width=0.95\textwidth]{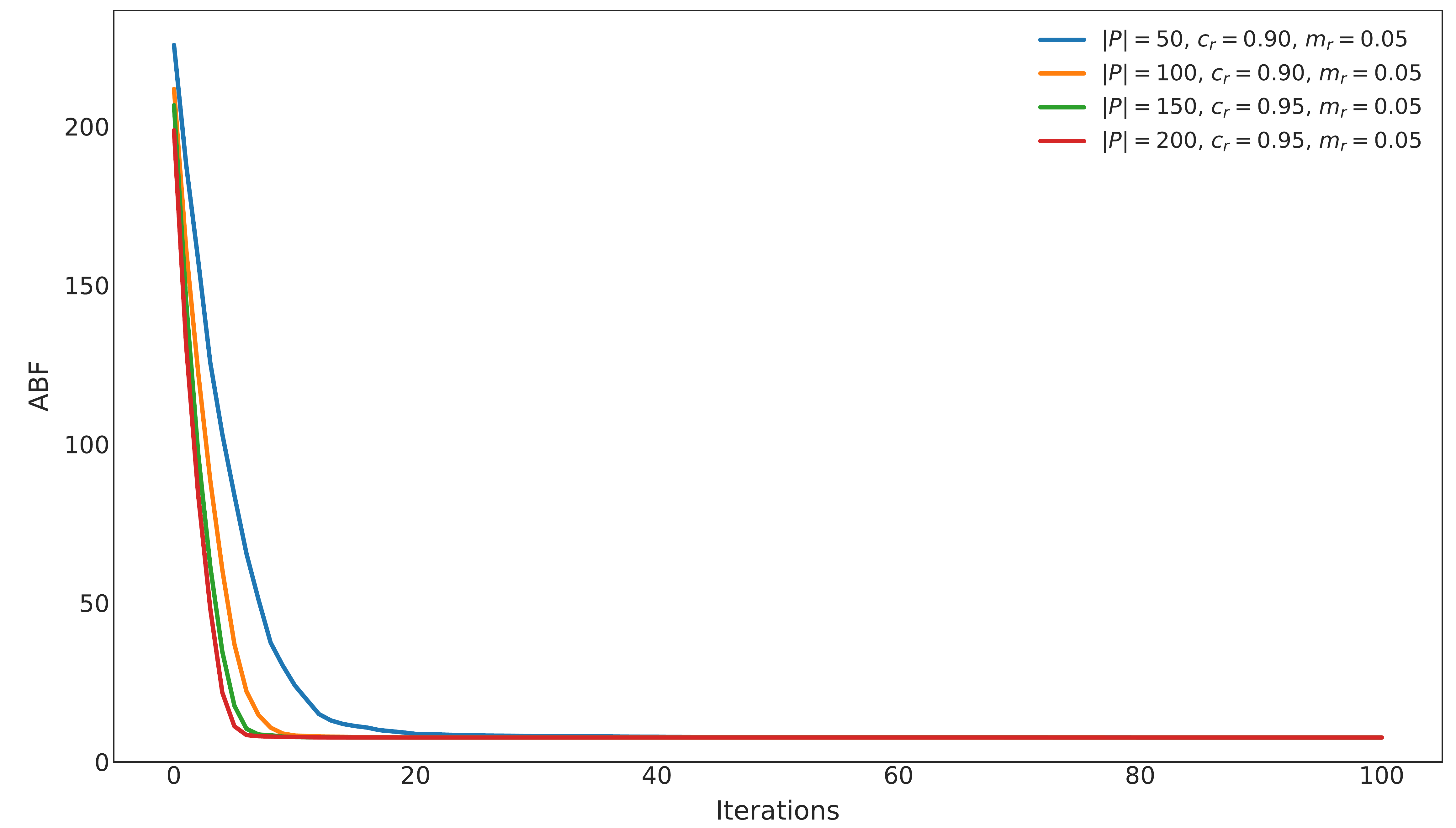}
  \caption{Comparison of the ABF achieved by GenHap with the best parameterizations found for each value of $|P|$ tested here.
The ABF was computed over the results of the optimization of instances characterized by $\#\text{SNPs} \in \{500, 1000, 5000\}$ and $f_\text{SNPs}=100$.}
%   \todo[inline]{da cosa si deduce la seq tech in questa figura? (io toglierei questa info dalla caption, eventualmente specificando a quali dati si riferisce) (D)}
  \label{fig:settingscomparison}
\end{figure}

% \begin{figure}[ht!]
%   \includegraphics[width=0.95\textwidth]{images/ABF.pdf}
%   \caption{Comparison of the ABF achieved by GenHap with the best parameterizations found for each value of $|P|$ tested here. The ABF was computed over the results of the optimization of $6$ instances with different $\#\text{SNPs}$ values as well as the sequencing technologies.}
%   \label{fig:settingscomparison}
% \end{figure}

As a first step, the performance of GenHap was evaluated to determine the best settings for the haplotype assembly problem.
We considered different instances for two sequencing technologies employed (i.e., Roche/454 and PacBio RS II), and we varied the settings of GenHap used throughout the optimization process, as follows:
\begin{itemize}
	\item  size of the population $|P| \in \{ 50, 100, 150, 200 \}$;
	\item  crossover rate $c_r \in \{ 0.8, 0.85, 0.9, 0.95 \}$;
	\item  mutation rate $m_r \in \{ 0.01, 0.05, 0.1, 0.15 \}$.
	%\item the size of the tournament selection strategy is fixed to $\kappa = \lceil 0.1 \cdot |P| \rceil $;
    %\item $T = 100$ iterations.
\end{itemize}
In all tests, the size of the tournament is fixed to $\kappa = 0.1 \cdot |P|$ and the maximum number of iterations is $T = 100$.
A total of $6$ different instances ($3$ resembling the Roche/454 sequencer and $3$ the PacBio RS II sequencer) were generated by considering $\#\text{SNPs} \in \{500, 1000, 5000\}$ and $f_\text{SNPs}=100$. %, \textcolor{red}{obtaining $6$ different instances}.
% Each different GenHap execution lasted $T = 100$ generations and was performed by varying one parameter at a time.

We varied one setting at a time, leading to $64$ different settings tested and a total number of $64\times 6 =384$ GenHap executions.
These tests 
%(data not shown) 
highlighted that, for each value of $|P|$, the best settings are:
\begin{enumerate}
 \item $|P|=50$, $p_c=0.9$, $p_m=0.05$;
 \item $|P|=100$, $p_c=0.9$, $p_m=0.05$;
 \item $|P|=150$, $p_c=0.95$, $p_m=0.05$;
 \item $|P|=200$, $p_c=0.95$, $p_m=0.05	$.
\end{enumerate}

Figure \ref{fig:settingscomparison} shows the comparison of the performance achieved by GenHap with the settings listed above, where the Average Best Fitness (ABF) was computed by taking into account, at each iteration, the fitness value of the best individuals over the $6$ optimization processes.
Even though all settings allowed GenHap to achieve almost the same final ABF value, we observe that the convergence speed increases with the size of the population. % as well as the running time required by GenHap.
%In order to choose the best settings, we analyzed the required running time concerning the $4$ tests listed above.
On the other hand, also the running time of GenHap increases with the size of the population. 
In particular, the executions lasted on average $1.41$ s, $2.33$ s, $3.52$ s, $4.95$ s with $|P| \in \{50, 100, 150, 200\}$,  respectively, running on one node of the Advanced Computing Center for Research and Education (ACCRE) at Vanderbilt University, Nashville, TN, USA.
The node is equipped with $2$ Intel\textsuperscript{\textregistered} Xeon\textsuperscript{\textregistered} E5-2630 v3 ($8$ cores at $2.40$ GHz) CPUs, $240$ GB of RAM and CentOS 7.0 operating system.
To perform the tests we exploited all $8$ physical cores of a single CPU.

%We also considered the accuracy achieved by GenHap as the average haplotype error rate (\textit{HE})  with respect to the ground truth \cite{andres2007}.
%The average \textit{HE} was $0.35$, $0.35$, $0.42$, $0.49$ for each $|P| \in \{50, 100, 150, 200\}$,  respectively.
Considering these preliminary results, we selected the parameter settings $|P|=100$, $c_r=0.9$, $m_r=0.05$, as the best trade-off between convergence speed (in terms of ABF) and running time.
% \todo[inline]{Visto che qui si fa riferimento al numero di errori rispetto al ground truth, andrebbe specificato il valore per ogni setting (P)}

\subsection*{Performance of GenHap}

\begin{table}[t]
\scriptsize
\centering
\caption{Comparison of GenHap and HapCol on the Roche/454 dataset with $\text{cov} \simeq 30\times$.
The performances were evaluated both in terms of \textit{HE} and  running time.
The N/A symbol denotes that HapCol was not able to complete the execution on all the $15$ instances.}
\label{tab:GenHapVSHapCol_r454}
\begin{tabular}{p{.8cm}p{1cm}p{1cm}|p{1cm}p{1cm}p{1cm}|p{1cm}p{1cm}p{1cm}}
\hline\hline
 &  &  & \multicolumn{3}{c|}{GenHap} & \multicolumn{3}{c}{HapCol} \\ \hline
$f_{\text{SNPs}}$ & $\text{cov}$ & $\#\text{SNPs}$ & Avg \textit{HE} & Std dev \textit{HE} & Avg Running Time {[}s{]} & Avg \textit{HE} & Std dev \textit{HE} & Avg Running Time {[}s{]} \\ \hline
\multirow{4}{*}{100} & \multirow{4}{*}{$\sim 30\times$}
  &   500 &  0.04 &  0.08 &  0.21 &  0.00 &  0.00 &  0.62 \\
& &  1000 &  0.09 &  0.08 &  0.36 &  0.00 &  0.00 &  1.20 \\
& &  5000 &  0.18 &  0.06 &  3.17 &  0.01 &  0.03 &  5.35 \\
& & 10000 &  2.50 &  5.52 & 10.33 &  6.55 & 16.38 & 10.23 \\ \hline
\multirow{4}{*}{200} & \multirow{4}{*}{$\sim 30\times$}
  &   500 &  0.09 &  0.14 &  0.34 &  0.00 &  0.00 &  0.50 \\
& &  1000 &  0.09 &  0.10 &  0.63 &  0.01 &  0.03 &  0.96 \\
& &  5000 &  3.61 &  3.43 &  6.07 &  0.38 &  0.78 &  4.90 \\
& & 10000 &  2.15 &  1.62 & 17.24 &    N/A &    N/A &    N/A \\
\hline\hline
\end{tabular}
\end{table}
\begin{figure}[t]
	\includegraphics[width=0.95\textwidth]{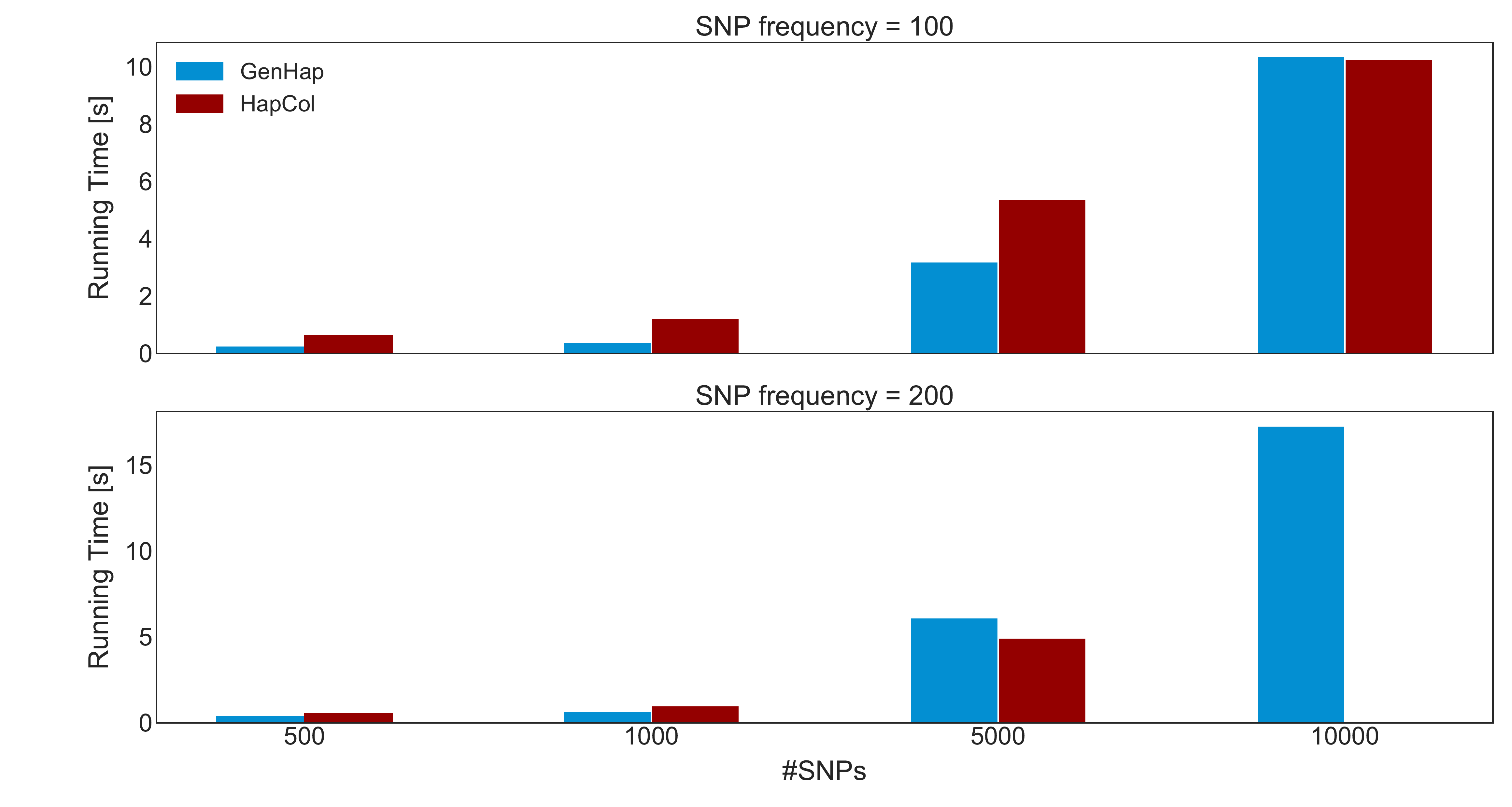}
	\caption{Comparison of the average running time required by GenHap (blue bars) and HapCol (red bars) computed over $15$ instances for each value of $\#\text{SNPs} \in \{500, 1000, 5000\}$ obtained with the Roche/454 sequencing technology, $\text{cov} \simeq 30\times$ and $f_{\text{SNPs}}=100$ (top) and $f_{\text{SNPs}}=200$ (bottom).
In the case of $f_{\text{SNPs}}=200$ and $\#\text{SNPs}=10000$, HapCol was not able to complete the execution on all the $15$ instances.}
\label{fig:timer454}
\end{figure}

% The performances of GenHap were compared against a state of the art tool, called HapCol \cite{pirola2015}, using the two datasets described above.
The performance achieved by GenHap was compared with those obtained by HapCol \cite{pirola2015}, which was shown to outperform the main available haplotyping approaches.
In particular, we exploited here a more recent version of HapCol capable of dealing with haplotype blocks \cite{Beretta170225}.
% \todo[inline]{immagino serva un rif bib per l'ultima versione di HapCol... (D)
% è stata aggiornata la versione nel repo, non c'è un paper associato (A)}
The same computational platform used for the setting analysis of GenHap was used to execute all the tests on the two synthetic datasets described above.
% \todo[inline]{Specificare che abbiamo usato la nuova versione di HapCol che gestisce gli haplotype block e sfrutta il pruning delle read. (Leo)}
% All tests on the two datasets described above were executed on the same computational platform used for the setting analysis of GenHap.
\begin{table}[t]
\scriptsize
\centering
\caption{Comparison of GenHap and HapCol on the PacBio RS II dataset with $\text{cov} \simeq 30\times$.
The performances were evaluated both in terms of \textit{HE} and running time.}
\label{tab:GenHapVSHapCol_PacBio}
\begin{tabular}{p{.8cm}p{1cm}p{1cm}|p{1cm}p{1cm}p{1cm}|p{1cm}p{1cm}p{1cm}}
\hline\hline
 &  &  & \multicolumn{3}{c|}{GenHap} & \multicolumn{3}{c}{HapCol} \\ \hline
$f_{\text{SNPs}}$ & $\text{cov}$ & $\#\text{SNPs}$ & Avg \textit{HE} & Std dev \textit{HE} & Avg Running Time {[}s{]} & Avg \textit{HE} & Std dev \textit{HE} & Avg Running Time {[}s{]} \\ \hline
\multirow{5}{1.5cm}{100} & \multirow{5}{*}{$\sim 30\times$}
  &   500 &  2.04 &  0.59 &  0.11 &  2.42 &  0.78 &  2.24 \\
& &  1000 &  1.27 &  0.51 &  0.19 &  1.20 &  0.61 &  1.89 \\
& &  5000 &  1.06 &  0.19 &  0.94 &  0.60 &  0.17 &  9.04 \\
& & 10000 &  0.96 &  0.19 &  2.50 &  0.43 &  0.11 & 15.51 \\
& & 20000 &  1.02 &  0.14 &  8.49 &  0.41 &  0.11 & 31.13 \\ \hline
\multirow{5}{1.5cm}{200} & \multirow{5}{*}{$\sim 30\times$}
  &   500 &  2.09 &  0.52 &  0.14 &  1.73 &  0.42 &  0.95 \\
& &  1000 &  1.70 &  0.24 &  0.22 &  1.09 &  0.41 &  1.84 \\
& &  5000 &  1.05 &  0.18 &  1.39 &  0.54 &  0.11 &  7.10 \\
& & 10000 &  1.13 &  0.18 &  4.09 &  0.51 &  0.17 & 14.13 \\
& & 20000 &  1.02 &  0.13 & 13.86 &  0.33 &  0.05 & 27.55 \\
\hline\hline
\end{tabular}
\end{table}
\begin{figure}[t!]
  \includegraphics[width=0.95\textwidth]{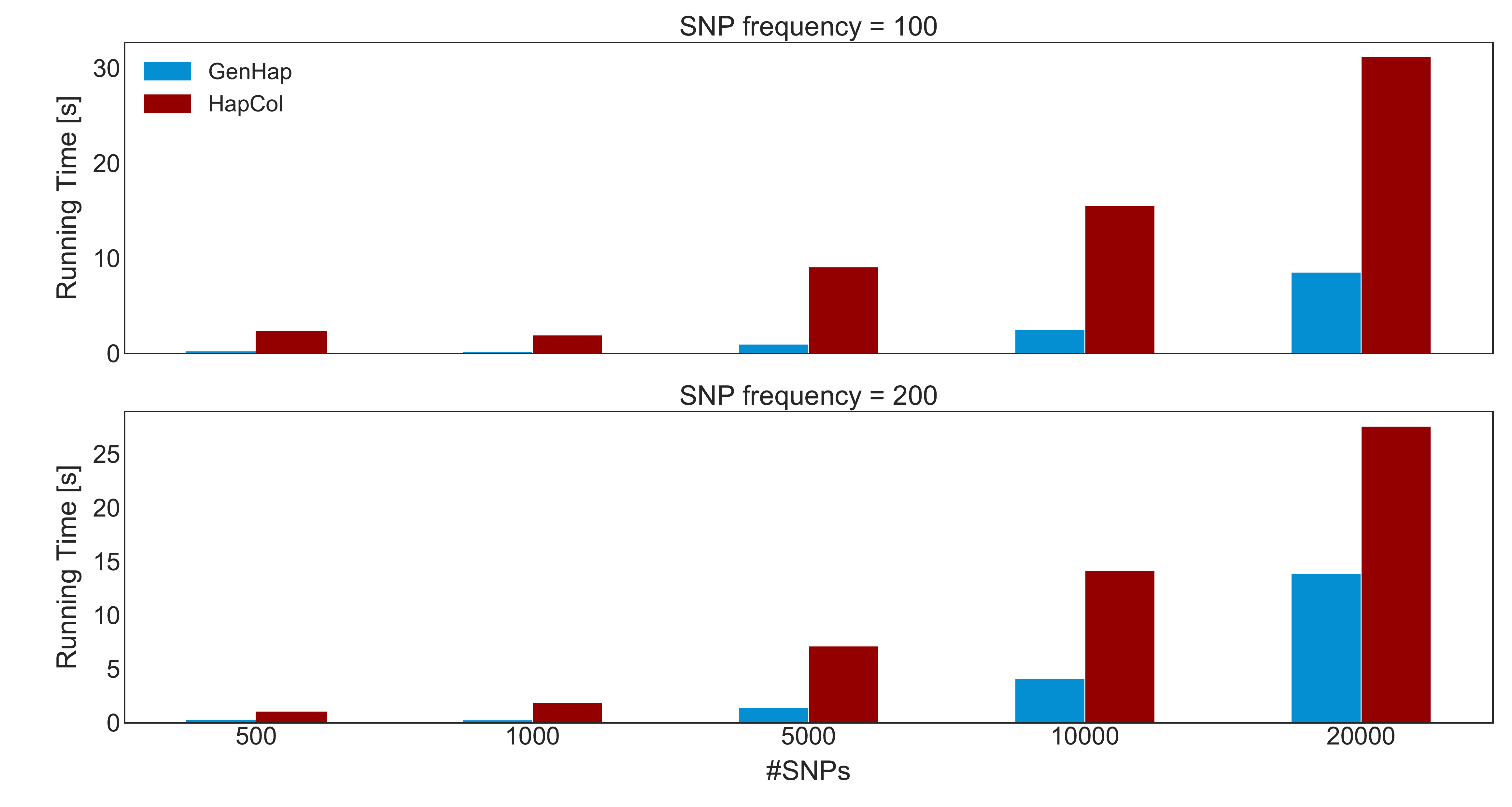}
  \caption{Comparison of the average running time required by GenHap (blue bars) and HapCol (red bars) computed over $15$ instances for each $\#\text{SNPs} \in \{500, 1000, 5000, 10000, 20000\}$ obtained with the PacBio RS II sequencing technology, $\text{cov} \simeq 30\times$, $f_{\text{SNPs}}=100$ (top) and $f_{\text{SNPs}}=200$ (bottom).}
  \label{fig:timePacBio}
\end{figure}

We stress the fact that GenHap was compared against HapCol only on the instances with $\text{cov} \simeq 30\times$, since HapCol is not capable of solving instances with higher coverage values (i.e., the algorithm execution halts when a column covered by more than $30$ reads is found).
% \textcolor{red}{As a matter of fact, in the case of sequencing data with $\text{cov} > 30\times$, a pre-processing step to reduce the coverage inside the blocks BLABLA is necessary before executing HapCoexecuting HapCol.l.}

Considering the two sequencing technologies, we generated $15$ different instances for each value of $\#\text{SNPs}$ and $f_\text{SNPs}$.
The performance was then evaluated by computing (\textit{i}) the average haplotype error rate (\textit{HE}), which represents the percentage of SNPs erroneously assigned with respect to the ground truth \cite{andres2007}, and (\textit{ii}) the average running time.

As shown in Table \ref{tab:GenHapVSHapCol_r454}, in the instances generated using the Roche/454 sequencing technology with $f_{\text{SNPs}} = 100$, both GenHap and HapCol reconstructed the two haplotypes, achieving an average \textit{HE} lower than $0.2\%$ with a negligible standard deviation in the case of $\#\text{SNPs} \in \{500, 1000, 5000\}$.
GenHap inferred the haplotypes characterized by $10000$ SNPs with an average \textit{HE} lower than $2.5\%$ and a standard deviation around $5\%$, while HapCol obtained an average \textit{HE} equal to $6.55\%$ with a standard deviation  around $16\%$.
For what concerns the running time, GenHap outperformed HapCol in all tests except in the case of $\#\text{SNPs}=10000$, as shown in  Figure \ref{fig:timer454}, being around $4\times$ faster in reconstructing the haplotypes.
In the case of $\#\text{SNPs}=10000$, the running times are comparable, but GenHap obtains a lower \textit{HE} than HapCol.
In the instances generated using $f_{\text{SNPs}} = 200$ and $\#\text{SNPs} \in \{500, 1000\}$, both GenHap and HapCol reconstructed the two haplotypes, achieving an average \textit{HE} lower than $0.1\%$ with a negligible standard deviation.
When $\#\text{SNPs} \in \{5000, 10000\}$ are taken into account, GenHap inferred the haplotype pairs with an average \textit{HE} lower than $3.65\%$ and a standard deviation lower than $3.5\%$.
Notice that HapCol was not able to complete the execution on all the $15$ instances characterized by $10000$ SNPs.
As in the case of instances with $f_{\text{SNPs}} = 100$, GenHap is faster than HapCol in all tests, except in the case of $\#\text{SNPs}=5000$.
%It is worth noting that the running time increases with $\#\text{SNPs}$ as also the number of reads increases according to Equation (\ref{eq:reads}), leading to a greater number $\Pi$  of sub-problems to be solved.

%We highlight that almost all instances with $f_{\text{SNPs}} = 100$ and $\#\text{SNPs} \in \{10000, 20000 \}$, as well as all instances with $f_{\text{SNPs}} = 200$, 
%We did not calculate the \textit{HE} for $\#\text{SNPs} \in \{10000, 20000 \}$ because almost all instances 
%were characterized by haplotype blocks \cite{daly2001}, making them unfeasible for the reconstruction of the full haplotypes.
%We highlight that almost all the instances related to this sequencing technology generated with $f_{\text{SNPs}} = 200$ were characterized by haplotype blocks, and for this reason we did not report the results.
%As a matter of fact, two haplotypes cannot be correctly reconstructed when haplotype blocks occur, as in the fragment matrix there exist at least two consecutive columns that are not covered by at least a common read.
%For this reason, we did not report the results achieved in these tests.
%\todo[inline]{sono un po' confusa, i risultati per SNP=10000 ci sono... (D) Ricordarsi di sistemare secondo i suggerimenti di Paolo su WhatsApp (D)}
% \todo[inline]{Spieghiamo meglio le limitazioni imposte dagli haplotype block. (L)}

%For what concerns the PacBio RS II sequencing dataset, thanks to the very long reads produced by this technology, we were able to test the two methods using $f_{\text{SNPs}} \in \{100, 200\}$ on  instances.
%\todo[inline]{all instances?}
For what concerns the PacBio RS II sequencing dataset, since this technology is characterized by a higher error rate with respect to the Roche/454 sequencer, both GenHap and HapCol reconstructed the two haplotypes with higher \textit{HE} values
%Since the synthetic dataset generated using this technology has a higher error rate with respect to those generated with the Roche/454 sequencer, both GenHap and HapCol reconstructed the two haplotypes with higher \textit{HE} values 
(see Table \ref{tab:GenHapVSHapCol_PacBio}).
Nonetheless, the average \textit{HE} value is lower than $2.5\%$ with a standard deviation lower than $1\%$ in all cases.
Figure \ref{fig:timePacBio} shows the running time required by GenHap and HapCol to reconstruct the haplotypes.
As in the case of the Roche/r454 dataset, the running time increases with $\#\text{SNPs}$, but GenHap always outperforms HapCol, achieving up to $20\times$ speed-up.

\begin{table}[t]
\scriptsize
\centering
\caption{Results obtained by GenHap on  the Roche/454 dataset with $\text{cov} \simeq 60\times$.
The performances were evaluated both in terms of \textit{HE} and running time.}
\label{tab:GenHap60_r454}
\begin{tabular}{p{1.3cm}p{1cm}p{1cm}|p{1cm}p{1cm}p{1cm}}
\hline\hline
 &  &  & \multicolumn{3}{c}{GenHap} \\ \hline
$f_{\text{SNPs}}$ & $\text{cov}$ & $\#\text{SNPs}$ & Avg \textit{HE} & Std dev \textit{HE} & Avg Running Time {[}s{]} \\ \hline
\multirow{4}{*}{100} & \multirow{4}{*}{$\sim 60\times$} 
  &   500 &  0.00 &  0.00 &  0.26 \\
& &  1000 &  0.05 &  0.05 &  0.54 \\
& &  5000 &  0.10 &  0.03 &  6.57 \\
& & 10000 &  0.15 &  0.03 & 21.13 \\\hline
\multirow{4}{*}{200} & \multirow{4}{*}{$\sim 60\times$} 
  &   500 &  0.00 &  0.00 &  0.37 \\
& &  1000 &  0.07 &  0.09 &  0.89 \\
& &  5000 &  1.13 &  1.72 & 11.17 \\
& & 10000 &  2.00 &  1.02 & 53.77 \\
\hline\hline
\end{tabular}
\end{table}
% \begin{figure}[t]
%   \includegraphics[width=0.95\textwidth]{images/HE_r454.pdf}
%   \caption{Bar diagram showing the \textit{HE} related to the results achieved by GenHap ($\text{cov} \simeq 60\times$ and $f_{\text{SNPs}}=100$) on $15$ instances generated for each $\#\text{SNPs} \in \{500, 1000, 5000, 10000\}$ with the Roche/454 sequencing technology.
% Black bars represent the standard deviation of \textit{HE}.}
%   \label{fig:HE_60x_r454}
% \end{figure}

Table \ref{tab:GenHap60_r454} lists the results obtained by GenHap on the instances of the Roche/454 dataset characterized by $\text{cov}\simeq 60\times$, $\#\text{SNPs} \in \{500, 1000, 5000, 10000\}$ and $f_{\text{SNPs}} \in \{100, 200\}$.
In all tests with $f_{\text{SNPs}} = 100$, GenHap was always able to infer the two haplotypes with high accuracy, indeed the average \textit{HE} values are always lower than $0.15\%$.
In the instances generated with $f_{\text{SNPs}} = 200$, GenHap reconstructed the haplotype pairs with an average \textit{HE} lower than $0.2\%$.
This interesting result shows that higher coverages can help during the reconstruction phase, allowing GenHap to infer more precise haplotypes.
% Figure \ref{fig:HE_60x_r454} allows to observe the \textit{HE} trend, whose value increases along with $\#\text{SNPs}$.
% \todo[inline]{ma serve davvero questa figura? (P)}

Regarding the PacBio RS II dataset, the achieved \textit{HE} is on average lower than $1.25\%$ with a standard deviation $\leq 0.4\%$ (see Table \ref{tab:GenHap60_PacBio}). 
In particular, the average \textit{HE} decreases when the value of $\#\text{SNPs}$ or the coverage increase, thus suggesting that higher cov values can considerably help in achieving a correct reconstruction of the two haplotypes.
On the contrary, the running time increases at most linearly with respect to the coverage (see Table \ref{tab:GenHap60_PacBio}).

As a first test on real sequencing data, we exploited a WIF input file codifying the SNPs of the chromosome $22$ generated from high-quality sequencing data  made publicly available by the GIAB Consortium.
This instance contains $\#\text{SNPs}\simeq27000$ and $\#\text{reads}\simeq80000$ with average and maximum coverages equal to $22$ and $25$, respectively.
In \cite{Beretta170225}, in order to down-sample the instances to the target maximum coverages of $30\times$ allowed by HapCol, the authors applied a greedy-based pruning strategy.
This procedure selects the reads characterized by high base-calling quality.
GenHap detected and inferred the $305$ different haplotype blocks in less than $10$ minutes, obtaining approximately an $87\%$ agreement with respect to the HapCol solution.
This agreement was calculated considering every SNP of both haplotypes in each block.

We tested GenHap also on the chromosome $22$ sequenced using the PacBio RS II technology (publicly available at \cite{pacbio54}).
This instance contains $\#\text{SNPs}\simeq28000$ and $\#\text{reads}\simeq140000$ with average and maximum coverages equal to $29$ and $565$, respectively.
GenHap reconstructed the two haplotypes in about $10$ minutes.
This result shows that GenHap is capable of dealing with instances characterized by high coverages,  avoiding pruning pre-processing steps.

% \todo[inline]{Aggiungere che GenHap riesce a simulare (Ivan: Attenzione che simulare non è corretto, meglio ricostruire gli aplotipi)) dati PacBio con $\#\text{SNPs}=28K$, 140K read, max cov=565 e avg=29, 1 blocco, in meno di 10 minuti. (A) (Ivan: i dati li ho presi qua: https://www.pacb.com/blog/data-release-54x-long-read-coverage-for/ Questa parte secondo è importante, anche mostrando qualche risultato un grafico o qualunque altra cosa perchè GenHap è l'unico tool che riesce a computare questi dati senza pruning riduzioni e trucchi assortiti) }

% \todo[inline]{Anche fig 8 secondo me è inutile (P) Anche secondo me Fig. 7 e 8 non sono indispensabili, visto che non viene fatto il confronto con HapCol e sono solo una rappresentazione grafica dei dati in tabella (D)}

%Finally, the results presented here show that GenHap always obtains high accuracy values, it is approximately $2\times$ faster than HapCol in the case of Roche/454 dataset and up to $7\times$ faster in the case of the PacBio RS II dataset.

\begin{table}[t]
\scriptsize
\centering
\caption{Results obtained by GenHap on the PacBio RS II dataset with $\text{cov} \simeq 60\times$.
The performances were evaluated both in terms of \textit{HE} and running time.}
\label{tab:GenHap60_PacBio}
\begin{tabular}{p{1.3cm}p{1cm}p{1cm}|p{1cm}p{1cm}p{1cm}}
\hline\hline
 &  &  & \multicolumn{3}{c}{GenHap} \\ \hline
$f_{\text{SNPs}}$ & $\text{cov}$ & $\#\text{SNPs}$ & Avg \textit{HE} & Std dev \textit{HE} & Avg Running Time {[}s{]} \\ \hline
\multirow{5}{1.5cm}{100} & \multirow{5}{*}{$\sim 60\times$}    
  &   500 &  1.22 &  0.36 &  0.17 \\
& &  1000 &  0.88 &  0.21 &  0.33 \\
& &  5000 &  0.56 &  0.10 &  1.81 \\
& & 10000 &  0.62 &  0.10 &  5.34 \\
& & 20000 &  0.60 &  0.07 & 17.14 \\ \hline
\multirow{5}{1.5cm}{200} & \multirow{5}{*}{$\sim 60\times$}
  &   500 &  1.22 &  0.37 &  0.22 \\
& &  1000 &  0.79 &  0.27 &  0.36 \\
& &  5000 &  0.53 &  0.09 &  3.26 \\
& & 10000 &  0.45 &  0.08 &  8.01 \\
& & 20000 &  0.49 &  0.05 & 27.15 \\
\hline\hline
\end{tabular}
\end{table}

% \begin{figure}[th!]
%   \includegraphics[width=0.95\textwidth]{images/HE_PacBio.pdf}
%   \caption{Bar diagram showing the \textit{HE} related to the results achieved by GenHap ($\text{cov} \simeq 60\times$, $f_{\text{SNPs}}=100$ (top) and $f_{\text{SNPs}}=200$ (bottom)) on 15 instances generated for each $\#\text{SNPs} \in \{500, 1000, 5000, 10000, 20000\}$ with the PacBio sequencing technology.
%   Black bars represent the standard deviation of \textit{HE}.}
%   \label{fig:HE_60x_PacBio}
% \end{figure}

% \todo[inline]{Commenti sui Risultati da estendere e discutere con maggiore criticità. L'Abstract andrà aggiornato di conseguenza. (L)}

\section*{Discussion and conclusions}

%The reconstruction of haplotypes represents a hot topic in Computational Biology and Bioinformatics. 
In this paper we presented GenHap, a novel computational method based on GAs to solve the haplotyping problem, which is one of the hot topics in Computational Biology and Bioinformatics. 
The performance of GenHap was evaluated by considering synthetic (yet realistic) read datasets resembling the outputs produced by the Roche/454 and PacBio RS II sequencers.
The solutions yielded by GenHap are accurate, independently of the number, frequency and coverage of SNPs in the input instances, and without any \textit{a priori} hypothesis about the sequencing error distribution in the reads.

In practice, our method was conceived to deal with data characterized by high-coverage and long reads, produced by recent sequencing techniques.
The read accuracy achieved by novel sequencing technologies, such as PacBio RS II and Oxford Nanopore MinION, may be useful for several practical applications.
In the case of SNP detection and haplotype phasing in human samples, besides read accuracy, a high-coverage is required to reduce possible errors due to few reads that convey conflicting information \cite{jain2016}.
In \cite{sims2014}, the authors argued that an average coverage higher than $30\times$ is the \textit{de facto} standard.
As a matter of fact, the first human genome that was sequenced using Illumina short-read technology showed that, although almost all homozygous SNPs are detected at a $15\times$ average coverage, an average depth of $33\times$ is required to detect the same proportion of heterozygous SNPs.
%\todo[inline]{quest'ultima frase e' da sistemare, although e while cozzano fra di loro (D)}

GenHap was implemented with a distributed strategy that exploits a Master-Slave computing paradigm in order to speed up the required computations.
We showed that GenHap is remarkably faster than HapCol \cite{pirola2015}, achieving approximately a $4\times$ speed-up in the case of Roche/454 instances, and up to $20\times$ speed-up in the case of the PacBio RS II dataset.
In order to keep the running time constant when the number of SNPs increases, the number of available cores should increase proportionally with $\#\text{SNPs}$.
%In order to evaluate the effectiveness of our approach, we run GenHap on synthetic (yet realistic) datasets.
%\todo[inline]{1) Hi-C data; 2) haplotype blocks; 3) poli-ploidity; 4) multi-objective. (A)}

Differently from the other state-of-the-art algorithms, GenHap was designed for taking into account datasets produced by the third-generation sequencing technologies, characterized by longer reads and higher coverages with respect to the previous generations.
As a matter of fact, the experimental findings show that GenHap works better with the datasets produced by third-generation sequencers.
%could therefore highly benefit from GenHap, thanks to its capability in solving large combinatorial problems.
%Our experimental results show that GenHap is able to achieve correct solutions in short running times. 
%In particular, it is able to solve large instances of the wMEC problem, yielding optimal solutions by means of a global search process, without any \textit{a priori} hypothesis about the sequencing error distribution in reads.
Although several approaches have been proposed in literature to solve the haplotyping problem \cite{patterson2015,pirola2015}, GenHap can be easily adapted to exploit Hi-C data characterized by very high-coverages (up to $90 \times$), in combination with other sequencing methods for long-range haplotype phasing \cite{ben2016}.
Moreover, GenHap can be also extended to compute haplotypes in organisms with different ploidity \cite{aguiar2013,berger2014}.
Worthy of notice, GenHap could be easily reformulated to consider a multi-objective fitness function (e.g., by exploiting an approach similar to NSGA-III \cite{deb2014}).
%Towards leveraging multi-objective optimization, GenHap is not based on classic single-objective optimization methods, such as Simulated Annealing, which can be difficult to extend for multi-objective optimizations.
In this context, a possible future extension of this work would consist in introducing other objectives in the fitness function, such as the methylation patterns of the different chromosomes \cite{guo2017} or the gene proximity in maps achieved through Chromosome Conformation Capture (3C) experiments \cite{merelli2013}.
As a final note, we would like to point out that there is currently a paucity of up-to-date real benchmarks regarding the latest sequencing technologies.
Therefore, collecting a reliable set of human genome sequencing data acquired with different technologies against the corresponding ground truth can be beneficial for the development of future methods.

\section*{List of abbreviations}
\textbf{3C:} Chromosome Conformation Capture
\textbf{ABF:} Average Best Fitness
\textbf{ACCRE:} Advanced Computing Center for Research and Education
\textbf{BAM:} Binary Alignment/Map
\textbf{CPU:} Central Processing Unit
\textbf{EDA:} Estimation of Distribution Algorithm
\textbf{GA:} Genetic Algorithm
\textbf{GeneSIM:} General Error-Model based SIMulator
\textbf{GIAB:} Genome in a Bottle
\textbf{HE:} Haplotype Error rate
\textbf{MEC:} Minimum Correction Error
\textbf{MPI:} Message Passing Interface
\textbf{NGS:} Next-Generation Sequencing
\textbf{PEATH:} Probabilistic Evolutionary Algorithm with Toggling for Haplotyping
\textbf{SAM:} Sequence Alignment/Map
\textbf{SNP:} Single Nucleotide Polymorphism
\textbf{VCF:} Variant Call Format
\textbf{WIF:} WhatsHap Input Format
\textbf{wMEC:} weighted Minimum Correction Error

\begin{backmatter}

\section*{Acknowledgments}
  This work was conducted in part using the resources of the Advanced Computing Center for Research and Education at Vanderbilt University, Nashville, TN, USA.
  
\section*{Availability of data and materials}
GenHap is a cross-platform software, i.e., it can be compiled and executed on the main Unix-like operating systems: GNU/Linux, and Apple Mac OS X.
GenHap is written in C++ and exploits a Message https://www.overleaf.com/dashPassing Interface (MPI) implementation.
GenHap's source files and binary executable files, as well as the datasets used during testing, are available on GitHub: https://github.com/andrea-tango/GenHap.

\section*{Author's contributions}
Conceived the idea: AT, MSN, IM.
Designed the code: AT, SS, LR.
Implemented the code: AT.
Performed the experiments: AT, SS, LR, IM.
Analyzed the data: AT, SS, LR.
Wrote the manuscript: AT, SS, LR, MSN, PC, IM, DB.
Critically read the manuscript and contributed to the discussion of the whole work: PL, GM.

\section*{Competing interests}
  The authors declare that they have no competing interests.

% \section*{Acknowledgements}
%   Text for this section \ldots
%%  After submission of the .TEX file,                     %%
%%  you will be prompted to submit your .BBL file.         %%

% if your bibliography is in bibtex format, use those commands:
\bibliographystyle{bmc-mathphys} % Style BST file (bmc-mathphys, vancouver, spbasic).
\bibliography{bmc_article}      % Bibliography file (usually '*.bib' )

\end{backmatter}
\end{document}